\documentclass[aps,prc,reprint,groupedaddress,nofootinbib]{revtex4-2}

\usepackage[utf8]{inputenc}
\usepackage{graphicx}
\usepackage{amsmath}

\begin{document}

	\newcommand*{\Tmax}{T_{0,\textrm{max}}}
	
	\title{Thermal photon production in Gubser inviscid relativistic fluid dynamics}
	\author{Jean-Fran\c{c}ois Paquet}
	\affiliation{Department of Physics and Astronomy, Vanderbilt University, Nashville TN 37240}
	\affiliation{Department of Mathematics, Vanderbilt University, Nashville TN 37240}
	\date{\today}
	
	\begin{abstract}
		The Gubser solution to inviscid relativistic fluid dynamics is used to examine the role of transverse expansion on the energy spectrum of photons radiated by quark-gluon plasma. Transverse flow is shown to be a modest effect on the energy spectrum of photons as a whole, despite its large effect on rare high-energy photons produced at low temperatures. An exact expression is derived for the volume of the plasma as a function of its temperature. A simple formula is obtained for the energy spectrum of high-energy thermal photons, which is used to relate the inverse slope $T_{\textrm{eff}}$ of the photon spectrum at energy $E$ to the maximum temperature of the plasma $T_0$, finding $T_{\textrm{eff}} \approx  T_0/(1+\frac{5}{2} \frac{T_0}{E})$.
	\end{abstract}

	\maketitle

	\section{Introduction}
	
	The plasma of quarks and gluons that permeated the Universe microseconds after the Big Bang can be recreated in collisions of large nuclei at ultrarelativistic energies. Unlike the quark-gluon plasma of the early Universe, which is generally approximated as homogenous in space and cooling down because of the Hubble expansion of the Universe, collider-produced quark-gluon plasma is evolving in flat spacetime and is highly asymmetric in space: it is approximated as invariant under boost in the spatial direction along the collision axis, and inhomogeneous in the plane transverse to the collision axis. The cooldown of quark-gluon plasma produced in nuclear collisions is initially driven by the plasma's expansion along the collision axis, which is later accelerated by the transverse expansion of the plasma. The cooldown of the plasma from the longitudinal expansion can be estimated from the so-called ``Bjorken'' solution to inviscid relativistic fluid dynamic~\cite{Bjorken:1982qr}, which neglects the effect of the transverse expansion and yields\footnote{Assuming a conformal equation of state, $P=\epsilon/3$, which is a good approximation for quark-gluon plasma at high temperatures ($T \gtrsim 400$~MeV).} the well-known time dependence for the energy density  $\epsilon(\tau)=\epsilon(\tau_0) \left( \tau_0/\tau \right)^{4/3}$, where $\tau=\sqrt{t^2-z^2}$ is the longitudinal proper time, with $z$ the spatial direction along the collision axis, $t$ the time, and $\tau_0$ the value of $\tau$ at some early time.
	
	An exact solution to relativistic fluid dynamics that can account for both the longitudinal and transverse expansion of the plasma was identified in Refs~\cite{Gubser:2010ze,Gubser:2010ui}. This ``Gubser solution'' can yield spacetime temperature profiles for the quark-gluon plasma that are at least semi-realistic, although the conformal equation of state $P=\epsilon/3$ used to obtain the solution leads to a faster cooldown than the known equation of state of strongly-coupled quark-gluon plasma~\cite{Borsanyi:2013bia, HotQCD:2014kol}. The radial transverse symmetry of the Gubser solution is most appropriate for head-on collisions of nuclei, even though it does not capture the finer scale structures and asymmetries which give a distinctive momentum anisotropy to the particles produced in high-energy nuclear collisions~\cite{Heinz:2013th}. Nevertheless, the Gubser solution is well suited to study the effect of the transverse expansion of the plasma in a simple analytical setting.\footnote{Because the Gubser solution is a simplified description of heavy-ion collisions, we will not attempt to make a distinction between ``quark-gluon plasma'', which assumes deconfined degrees of freedom, and the ``hadronic plasma'' resulting from the reconfinement of quark-gluon plasma. Because of its conformal equation of state, the Gubser solution cannot describe appropriately the hadronic plasma.}
	
	In this work, we use the Gubser solution to study the radiation of high-energy photons from the hot and dense quark-gluon plasma.
	Unlike its early Universe counterpart, the quark-gluon plasma produced in nuclear collisions is largely transparent to high-energy electromagnetic radiation as a  consequence of the smaller size of the plasma, which is shorter than the photon's mean free path.
	Most photons measured in nuclear collisions are not radiated from the quark-gluon plasma; they are dominated by photonic decays of copiously produced hadrons (e.g. $\pi_0 \to \gamma \gamma$).
	However, for low photon energies ($E \lesssim 5$~GeV), there is a measurable signal~\cite{PHENIX:2014nkk,ALICE:2015xmh,STAR:2016use,PHENIX:2022rsx,PHENIX:2022qfp} that comes from neither hadronic decays nor prompt photons, and that is suggestive of blackbody photons radiated by the quark-gluon plasma~\cite{David:2019wpt,Paquet:2015lta,Kim:2016ylr,Dasgupta:2018pjm,Garcia-Montero:2019kjk,Monnai:2022hfs,Gale:2021emg,Chatterjee:2017akg, Holopainen:2011pd,Chatterjee:2013naa} . In particular, the measured photon ``excess'' has a distinctive exponential dependence on the photon energy~\cite{PHENIX:2014nkk,ALICE:2015xmh,STAR:2016use,PHENIX:2022rsx,PHENIX:2022qfp}. The inverse slope of the photon energy spectrum is often referred to as the ``effective temperature'' $T_{\textrm{eff}}$, because it is an indirect measure of the plasma's temperature, averaged over the plasma's complex temperature profile and further affected by the Doppler shift from the plasma's transverse expansion. The Gubser solution provides a simple semi-realistic setting to study the effect of the transverse expansion on the photon energy spectrum and on the inverse slope $T_{\textrm{eff}}$.

	\section{Gubser solution to inviscid relativistic fluid dynamics}
	
	The equations of inviscid relativistic fluid dynamics for the temperature field $T(X)$ and the flow velocity field $u^\mu(X)$ are given by
	\begin{align}
		u^\mu \partial_\mu \ln T &= -c_s^2(T) \partial_\mu u^\mu \\
		u^\nu \partial_\nu u^\mu &= \left[ g^{\mu\nu}+u^\mu u^\nu \right] \partial_\nu \ln T
	\end{align}
	with $g^{\mu\nu}=\textrm{diag}(-1,+1,+1,+1)$, $c_s(T)$ the speed of sound of the plasma and $X$ the spacetime position four-vector.
	
\begin{figure}[tbh]
		\centering
		\includegraphics[width=0.8\linewidth]{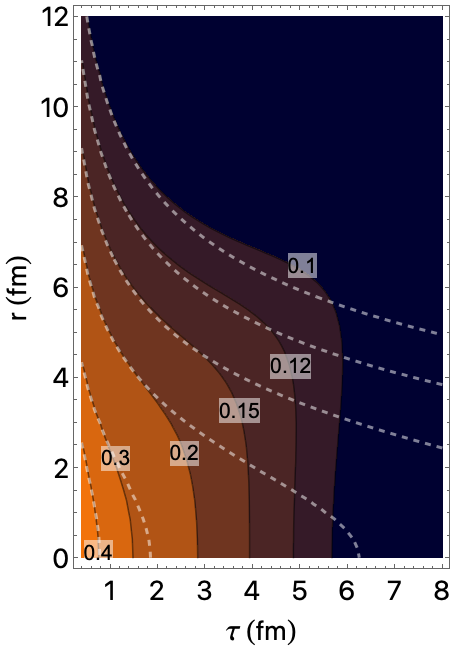}
		\caption{Temperature profile given by the Gubser solution (Eq.~\ref{eq:T_Gubser}) to inviscid relativistic fluid dynamics with parameters $T_0=0.5$~GeV, $q=0.25$~fm$^{-1}$ and $\tau_0=0.4$~fm. The labels of the temperature contours are in GeV. The white dashed lines are the contours obtained when the transverse expansion is neglected (Eq.~\ref{eq:Bjorken}), showing that the effect of the transverse expansion leads to a significantly faster cooldown, and smaller volumes of plasma (see Eq.~\ref{eq:dVdT_def}) between the temperature contours.}
		\label{fig:T_contour}
	\end{figure}

	For a conformal fluid ($c_s^2=1/3$) that is invariant under boosts in the direction of the collision axis and cylindrically symmetric in the plane transverse to the collision axis, there exists a family of exact ``Gubser'' solutions~\cite{Gubser:2010ze,Gubser:2010ui,Marrochio:2013wla}  given by
	\begin{equation}
		T(\tau,r)=\frac{\hat{T}_0 (2 q \tau)^{2/3}}{\tau \left(1+2 q^2 \left(\tau^2+r^2\right)+q^4 \left(\tau^2-r^2\right)^2\right)^{1/3}}
		\label{eq:T_Gubser}
	\end{equation}
	and
	\begin{equation}
		u^r(\tau,r)=\frac{2 r q^2 \tau}{\sqrt{1+2 q^2 \left(\tau^2+r^2\right)+q^4 \left(\tau^2-r^2\right)^2}} 
		\label{eq:flow_Gubser}
	\end{equation}
	where we use the coordinates ($\tau=\sqrt{t^2-z^2},r=\sqrt{x^2+y^2},\phi=\arctan(y/x),\eta_s=\text{arctanh}(z/t)$) defined such that $t=\tau \cosh(\eta_s)$, $x=r \cos(\phi)$, $y=r \sin(\phi)$ and $z=\tau \sinh(\eta_s)$.
	The normalization factor $\hat{T}_0$ is defined by
	\begin{equation}
		\hat{T}_0=T_0 \tau_0 \left( \frac{1+q^2 \tau_0^2}{2 q \tau_0} \right)^{2/3}
	\end{equation}
	such that $T(\tau_0,r=0)=T_0$. With these definitions, if we assume that the plasma is formed at time $\tau=\tau_0$, its maximum temperature is given by $T_0$, at the center of the plasma ($r=0$). Varying the parameter $q$ provides an ensemble of solutions. Because we want to mimic the plasma produced in ultrarelativistic nuclear collisions, we generally want to choose a relatively small value for $q$, as already pointed out in Refs.~\cite{Gubser:2010ze,Gubser:2010ui}. Using $T_0=0.5$~GeV, $q=0.25$~fm$^{-1}$ and $\tau_0=0.4$~fm, we get the temperature profile shown in Figure~\ref{fig:T_contour}.
	
	The radial width of the initial temperature profile is of order $1/q$. It can be seen from Eq.~\ref{eq:T_Gubser} that, as long as $q \tau$ and $q r$ are smaller than one, the temperature does not deviate significantly from the Bjorken-like solution~\cite{Bjorken:1982qr} 
	\begin{equation}
	\frac{T(\tau,r)}{T(\tau_0,r)}\approx \left(\frac{\tau_0}{\tau}\right)^{1/3} .
	\label{eq:Bjorken}
	\end{equation}
	This is a consequence of the transverse flow velocity being small at early time compared to the longitudinal flow velocity, and thus the cool down being dominated by the Bjorken-like longitudinal expansion. We show the result of Eq.~\ref{eq:Bjorken} in Figure~\ref{fig:T_contour} to emphasize how much faster the Gubser temperature profile cools down from the effect of transverse flow (filled contours), as opposed to the effect of the longitudinal cool down only (dashed lines).	
	
	\section{Photon production}
	
	In this inviscid hydrodynamics setting, we assume that the relaxation time of the plasma is much shorter than the characteristic length and time scales of the evolution (e.g. inverse gradients), such that the medium is always locally close to equilibrium, and entropy production is neglected. We correspondingly assume that the production of photons at each spacetime point is given by the equilibrium thermal emission rate of quark-gluon plasma, $k d \Gamma_{\gamma}/d^3 \mathbf{k}$~\cite{Kapusta:1991qp, Aurenche:1998nw, Arnold:2001ms,Ghiglieri:2013gia,Gale:2014dfa,Hidaka:2015ima,Ghiglieri:2016tvj,Ce:2022fot, Jackson:2019yao} for photons of momentum $\mathbf{k}$. Because absorption is minimal, the photon energy spectrum is obtained by the convolution of the thermal emission rate with the temperature and flow velocity profile:
	\begin{equation}
		k \frac{d^3 N}{d^3 \mathbf{k}} \!\! = \!\! \int_{\tau_0}^{\infty} \!\!\!\!\!\! d\tau \int_0^{\infty} \!\!\!\!\!\! dr \int_{-\infty}^{\infty} \!\!\!\!\!\!  d\eta_s \tau 2 \pi r  \left[ k \frac{d \Gamma_{\gamma}(K\cdot u,T)}{d^3 \mathbf{k}} \right] \Theta(T>T_{\textrm{min}})
		\label{eq:thermal_photon_spectrum}
	\end{equation}
	where both $T$ and $u^\mu$ are functions of $\tau$ and $r$. The theta function $\Theta(T>T_{\textrm{min}})$ cuts off photon emission at a minimum temperature $T_{\textrm{min}}$ to account for the breakdown of hydrodynamics and the reduction of photon emission after the quark-gluon plasma recombines into hadrons. It is typical that values between 100 and 150~MeV are used for $T_{\textrm{min}}$ when studying photon production in heavy-ion collisions~\cite{Gale:2021emg,Schafer:2021slz,Kim:2016ylr,Garcia-Montero:2019kjk,Monnai:2022hfs}.
	
	Using the quark-gluon plasma thermal photon emission rate from Ref.~\cite{Arnold:2001ms} and Equations~\ref{eq:T_Gubser} and \ref{eq:flow_Gubser} for the temperature and flow velocity profile, we can evaluate numerically the photon emission from the plasma.
	We decompose the photon momentum $K^\mu$ in the usual cylindrical-hyperbolic coordinates with the rapidity $y_M=\ln[(K^t-K^z)/(K^t+K^z)]/2$ characterizing the momentum along the collision axis, and $k_T$ and $\phi$ used in the transverse plane. We focus on photons with momentum perpendicular to the collision axis, $y_M=0$, in which case $E=k_T$ is the energy of the photons. We compute the photon energy spectrum per unit momentum rapidity, which is finite even for a plasma with an infinite extent in the longitudinal direction: $\left. 1/(2\pi E) dN/dE dy_M \right|_{y_M=0}$.
	
	\begin{figure}[tbh]
		\centering
		\includegraphics[width=\linewidth]{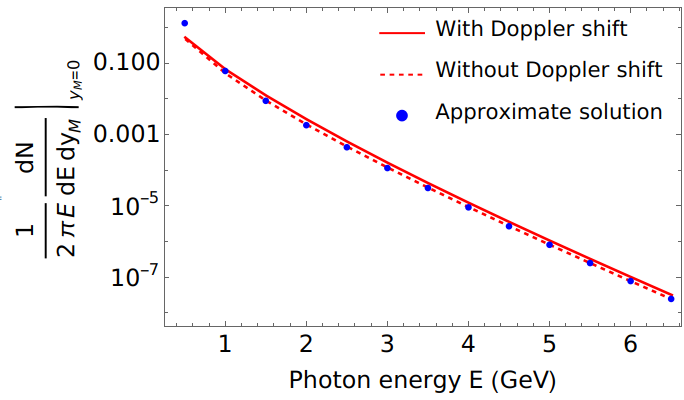}
		\caption{Photon energy spectrum resulting from the temperature profile shown in Figure~\ref{fig:T_contour}, computed with Eq.~\ref{eq:thermal_photon_spectrum} using the quark-gluon plasma photon emission rates from Ref.~\cite{Arnold:2001ms} and $T_{\textrm{min}}=100$~MeV. The solid line includes the effect of the Doppler shift from the flow velocity, while the dashed line neglects it. The blue dots show the approximate solution of Eq.~\ref{eq:thermal_photon_spectrum_gubser_approx} discussed below.}
		\label{fig:spectrum_with_without_flow}
	\end{figure}

\begin{figure}[tbh]
		\centering
		\includegraphics[width=\linewidth]{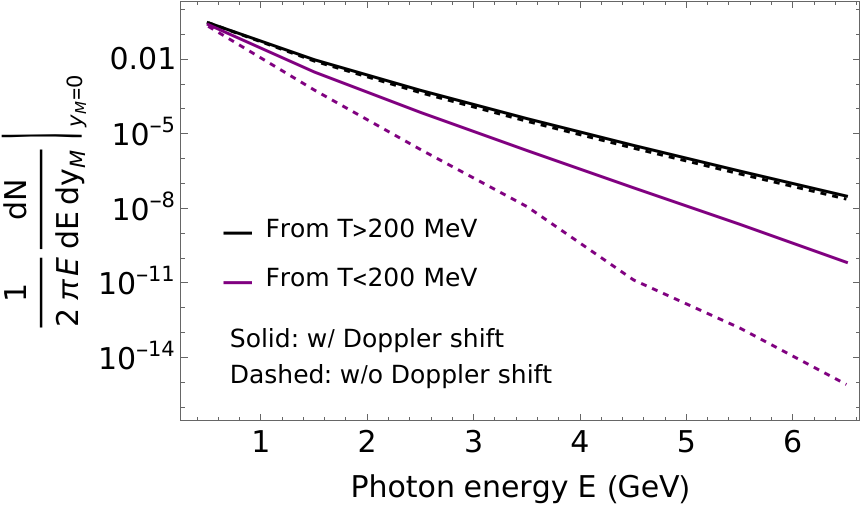}
		\caption{Same as Figure~\ref{fig:spectrum_with_without_flow}, with the contribution to the photon energy spectrum divided into the contribution from $T>200$~MeV (black lines) and $200>T>100$~MeV (purple lines).}
		\label{fig:spectrum_with_without_flow_breakdown}
	\end{figure}

	To quantify the effect of the transverse flow velocity, we compute the photon energy spectrum with and without the Doppler shift from the transverse flow; to turn off the Doppler shift, we set $u^r=0$ in $K\cdot u$ entering in the rate in Eq.~\ref{eq:thermal_photon_spectrum}. The result is  shown in Figure~\ref{fig:spectrum_with_without_flow}. 
	The effect of the transverse flow is moderate, less than a 30\% change in the spectrum for most of the energy range, which appears to conflict with the expected significance of the Doppler shift on the produced photons.  To understand this result, we separate the contribution of photons originating from temperatures above and below 200~MeV, and show the results in  Figure~\ref{fig:spectrum_with_without_flow_breakdown}. The Doppler shift on high-energy photons produced at low temperatures (bottom two lines) is large, enhancing the spectrum by three orders of magnitude for 4~GeV photons and five orders of magnitude for 6~GeV photons; this is similar to the effect seen in Ref.~\cite{Paquet:2016ime} when the Doppler shift is neglected in a state-of-the-art hydrodynamic simulation of heavy-ion collisions. Nevertheless, the effect on the overall spectrum is small, because very few 4~GeV photons are produced at low temperatures compared to the ones produced at high temperatures ($T>200$~MeV).
	As for the low-energy photon spectrum, photons produced at low temperatures do have a large contribution, but the effect of the Doppler shift is much smaller on these photons.	
	Thus, there is a clear distinction between the ``local'' effect of the Doppler shift, which can shift the photon spectrum by orders of magnitude, and the ``global'' effect of the Doppler shift once photons from all parts of the plasma are summed.

	If the transverse flow velocity is neglected, the photon energy spectrum can be written in a much simpler form~\cite{Shuryak:1978ij}:
	\begin{multline}
		\left. \frac{1}{2\pi E} \frac{d N}{d E dy_M} \right|_{y_M=0} \!\! = \\
		\!\!  \int_{T_{\textrm{min}}}^{T_0} \!\!\!\! dT \int_{-\infty}^{\infty} \!\!\!\!\!\!  d\eta_s \frac{d V_\perp}{d T}  \left[ k \frac{d \Gamma_{\gamma}(E \cosh(\eta_s),T)}{d^3 \mathbf{k}} \right] 
		\label{eq:thermal_photon_spectrum_no_flow}
	\end{multline}
	with the transverse volume per unit temperature given by
	\begin{equation}
		\frac{d V_\perp}{d T} = \int_{\tau_0}^{\infty} \!\!\!\!\!\! d\tau \int_0^{\infty} \!\!\!\!\!\! dr \tau 2 \pi r \delta(T-T(\tau,r)).
		\label{eq:dVdT_def}
	\end{equation}
	
	For the Gubser temperature profile, the transverse volume per unit temperature can be evaluated analytically. Defining $v=q \tau_0$, the result is 
	\begin{equation}
		\frac{d V_\perp}{d T} = \frac{\pi T_0^3 \tau_0^4 \left(1+v^2\right)^2}{2 v^3 T^4} F\left(\frac{T}{T_0},v\right)		\label{eq:dVdT_exact}
	\end{equation}
	with
	\begin{equation}
		F=
		\begin{cases}
			\arcsin(\beta \left(\frac{\delta}{v}\right)^{3/2})-\arcsin(\beta) & \text{if } \frac{T}{T_0}>\frac{\left[ v \left(1+v^2\right)^2 \right]^{1/3} }{2^{2/3}}\\
			\arccos(\beta) & \text{otherwise}
		\end{cases}  
		\label{eq:dVdT_exact_F_fct}
	\end{equation}
	where
	\begin{equation}
		\beta=\frac{2 v (T/T_0)^{3/2}}{1+v^2}; \; \delta \left(\delta^2+1\right)^2=\frac{v\left(1+v^2\right)^2}{\left(T/T_0\right)^3}; \; \delta>0.
	\end{equation}

	At high temperature, and assuming $v=q \tau_0 \ll 1$, which is the case for the parameters used to obtain the temperature profile in Figure~\ref{fig:T_contour}, Eq.~\ref{eq:dVdT_exact} can be approximated by:
	\begin{equation}
		\frac{d V_\perp}{d T} \approx
		\frac{\pi T_0^3 \tau_0^4 \left(1+v^2\right)^2}{2 v^3 T^4} \left[ 2 v \left(\frac{T_0^3}{T^3}-\left(\frac{T}{T_0}\right)^{3/2}\right) \right]
		\label{eq:dVdT_approx_highT}
	\end{equation}
	
	At low temperatures, Eq.~\ref{eq:dVdT_exact} takes the simple form
	\begin{equation}
	\frac{d V_\perp}{d T} \approx
	\frac{\pi T_0^3 \tau_0^4 \left(1+v^2\right)^2}{2 v^3 T^4}  \frac{\pi}{2} \; .
		\label{eq:dVdT_approx_lowT}
	\end{equation}

\begin{figure}[tb]
		\centering
		\includegraphics[width=\linewidth]{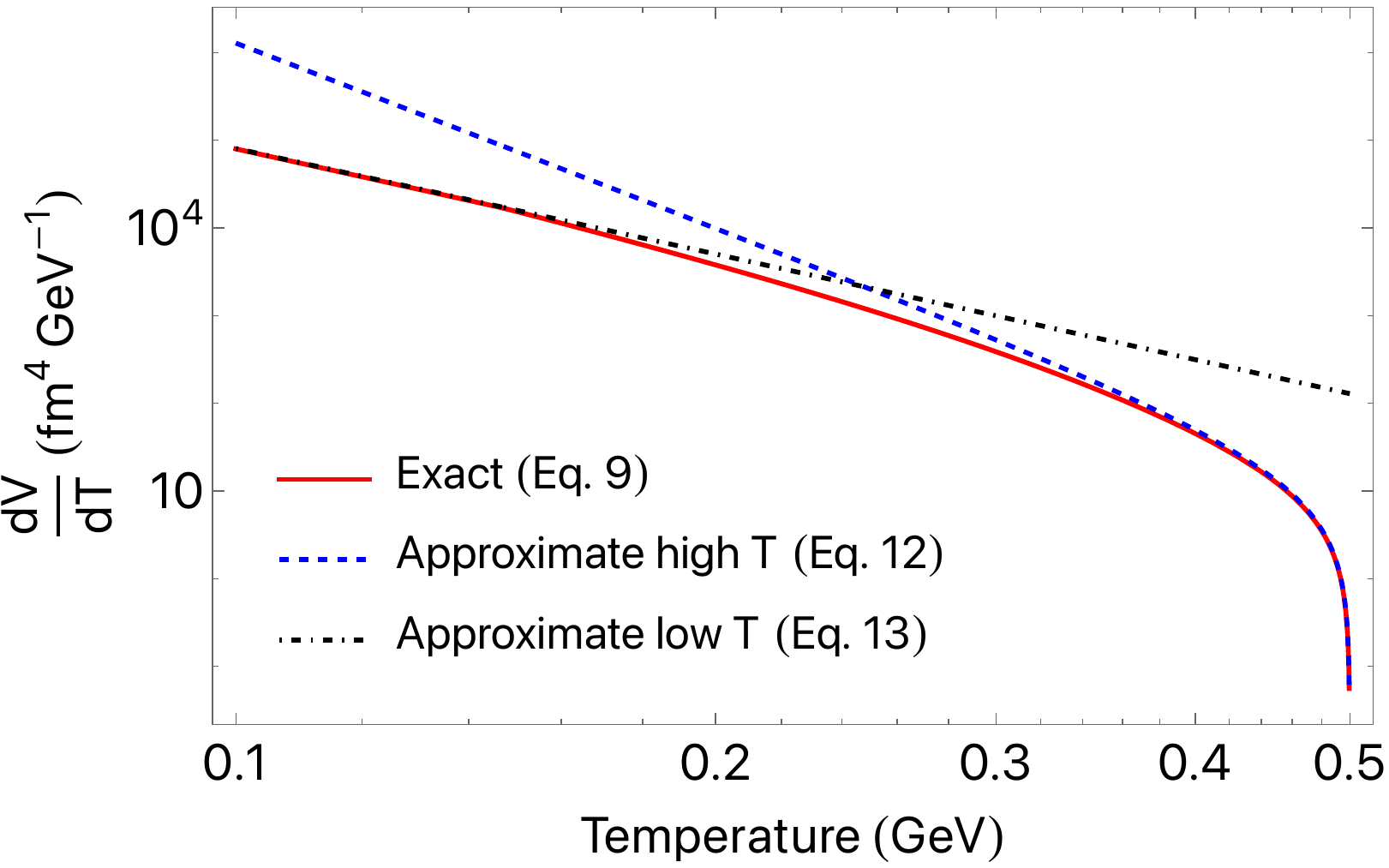}
		\caption{Transverse volume of plasma as a function of the temperature for the profile shown in Figure~\ref{fig:T_contour} given by the exact Eq.~\ref{eq:dVdT_exact} (red line), compared to the approximate Eq.~\ref{eq:dVdT_approx_highT} (dashed blue line) and Eq.~\ref{eq:dVdT_approx_lowT} (dash-dotted black line).}
		\label{fig:volume}
	\end{figure}

	Equations~\ref{eq:dVdT_approx_highT} and \ref{eq:dVdT_approx_lowT} are compared with the exact Eq.~\ref{eq:dVdT_exact} in Figure~\ref{fig:volume}. The  $T^{-4}$ dependence of the volume at low temperature departs considerably from the $T^{-7}$ found at high temperature~\cite{Shuryak:1978ij,McLerran:1984ay} when the transverse expansion is negligible; this can also be seen in Figure~\ref{fig:T_contour} by comparing the different temperature contours with the corresponding dashed lines representing the result without transverse expansion: $d V_\perp/d T \propto T^{-7}$  results in much larger volume of plasma at low temperature (dashed lines) than a $T^{-4}$ dependence (filled contours).
	
	Using the high-temperature expansion (Eq.~\ref{eq:dVdT_approx_highT}) and assuming that the dominant temperature and energy dependence of the photon emission rate is $T^2 \exp(-E/T)$, which is a good approximation for the quark-gluon plasma rate~\cite{Arnold:2001ms}, Eq.~\ref{eq:thermal_photon_spectrum_no_flow} can be integrated to yield the following simple expression for the production of photons in inviscid Gubser hydrodynamics:
	\begin{multline}
		\left. \frac{1}{2\pi E} \frac{d N}{d E dy_M} \right|_{y_M=0} \!\! \approx \!\!  \frac{9 \pi ^{3/2} \tau_0^2  \left(1+q^2 \tau_0^2\right)^2}{\sqrt{2} q^2} \left(\frac{T_0}{E}\right)^{5/2} \\ \left[ k \frac{d \Gamma_{\gamma}(E,T_0)}{d^3 \mathbf{k}} \right]
		\label{eq:thermal_photon_spectrum_gubser_approx}
	\end{multline}
	A comparison of Eq.~\ref{eq:thermal_photon_spectrum_gubser_approx} with the exact calculation is shown in Figure~\ref{fig:spectrum_with_without_flow}, again for the hydrodynamics profile from Figure~\ref{fig:T_contour}. The formula can be seen to work very well for high-energy photons ($E \gtrsim 2$~GeV), which are dominantly produced at high temperatures where Eq.~\ref{eq:dVdT_approx_highT} is a good approximation of the volume.

	Equation~\ref{eq:thermal_photon_spectrum_gubser_approx} is not exponential in the photon energy because of the $(T_0/E)^{5/2}$ term, and because the thermal photon emission rate itself is not exactly exponential. If one were to attempt to fit the slope\footnote{The \emph{slope} of the photon spectrum is the quantity that is fitted to extract an ``effective temperature''. Fitting the spectrum itself and extracting an effective temperature using 	\begin{equation}
		(\textrm{constant})-\frac{E}{T_{\textrm{eff}}^\prime} \approx \frac{5}{2} \ln \left( \frac{T_0}{E} \right) + \ln \left[ k \frac{d \Gamma_{\gamma}(E,T_0)}{d^3 \mathbf{k}} \right]
	\end{equation}
	would yield an effective temperature that differs from the traditional definition of the inverse slope.
	} of photon energy spectrum with an $\exp(-E/T_{\textrm{eff}})$ function, one would find
	\begin{equation}
		\frac{d}{d E} \left[ -\frac{E}{T_{\textrm{eff}}} \right] \approx \frac{d}{d E} \left[ \frac{5}{2} \ln \left( \frac{T_0}{E} \right) + \ln \left[ k \frac{d \Gamma_{\gamma}(E,T_0)}{d^3 \mathbf{k}} \right] \right]
		\label{eq:Teff_def}
	\end{equation}
	which, neglecting the non-exponential corrections to the rate,
    implies
	\begin{equation}
		T_{\textrm{eff}}\approx \frac{T_0}{1+\frac{5}{2} \frac{T_0}{E}}.
		\label{eq:Teff}
	\end{equation}
	
		\begin{figure}[tb]
		\centering
		\includegraphics[width=\linewidth]{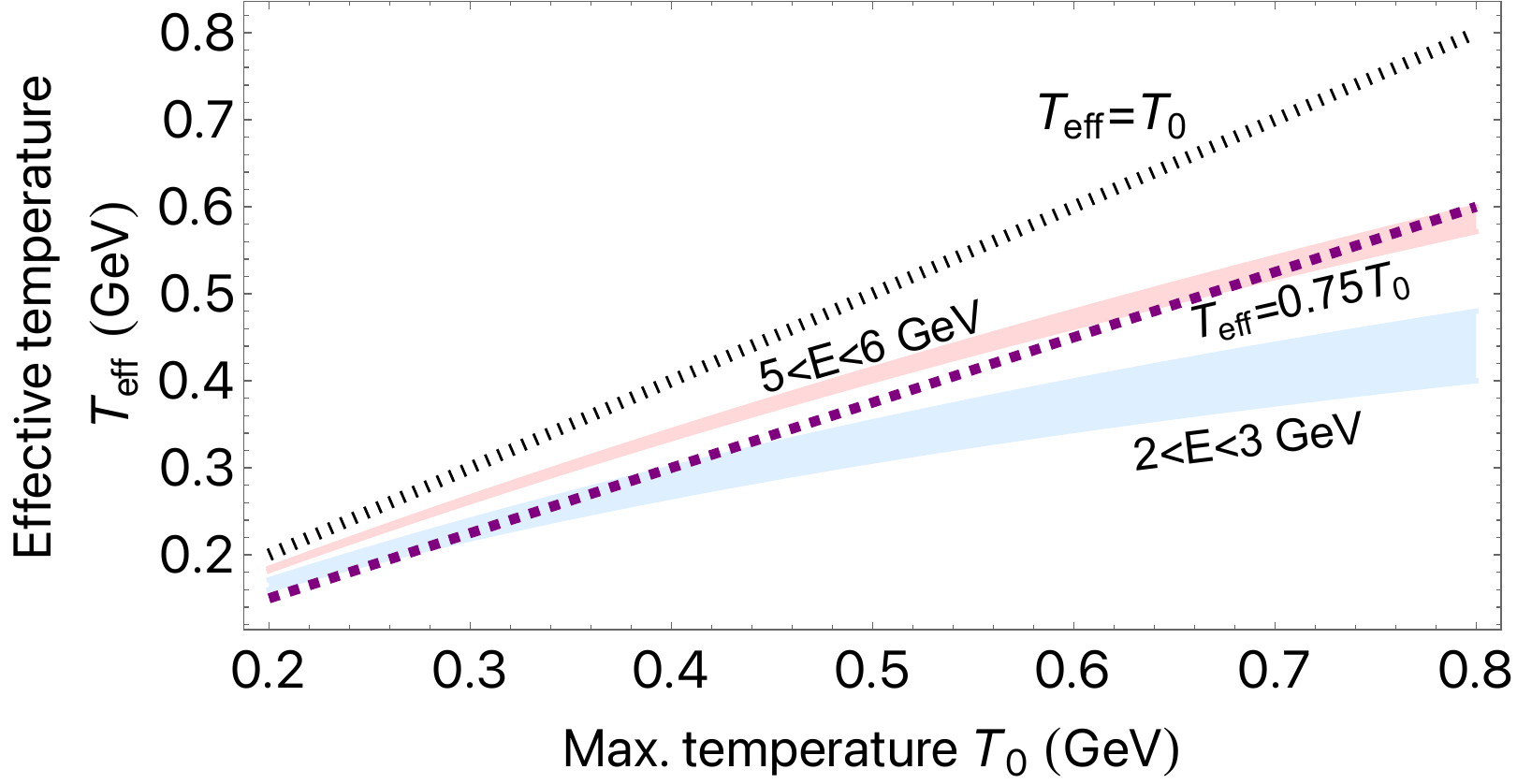}
		\caption{Effective temperature $T_{\textrm{eff}}$ given by Eq.~\ref{eq:Teff}, extracted from the photon energy spectrum (see discussion above Eq.~\ref{eq:Teff_def}), as a function of the actual maximum temperature $T_0$ of the plasma, for two different ranges of photon energy. The slopes $T_{\textrm{eff}} = T_0$ and $T_{\textrm{eff}} = 0.75 T_0$ are plotted to guide the eye.}
		\label{fig:Teff_vs_T0}
	\end{figure}

In Figure~\ref{fig:Teff_vs_T0}, the effective temperature $T_{\textrm{eff}}$ is  plotted as a function of the maximum plasma temperature $T_0$ for two different photon energy ranges relevant in heavy-ion collisions. The line $T_{\textrm{eff}} \approx 0.75 T_0$ is plotted to guide the eye, although the dependence of the effective temperature on $T_0/E$ is relatively strong and non-linear for lower-energy photons. Attempting to measure high plasma temperatures with somewhat low-energy photons would lead to particularly large differences between $T_{\textrm{eff}}$ and $T_0$: for example, with $T_0=0.5$~GeV and $E=2$~GeV, we find $T_{\textrm{eff}}=0.307$~GeV.

	\section{Summary}
	
Quark-gluon plasma produced in ultrarelativistic nuclear collisions cools down by a combination of longitudinal and later transverse expansion. The Gubser solution discussed in this work can be used to better understand the combined effect of these expansions. Photons are expected to undergo a considerable blueshift due to the transverse expansion, which can be seen in Figure~\ref{fig:spectrum_with_without_flow_breakdown} for a Gubser hydrodynamic profile. 
However, this effect is concentrated on higher-energy photons produced at low temperatures.
The \emph{overall} effect of this Doppler shift is much smaller, because these Doppler-shifted photons are a subdominant source of high-energy photons.

Heavy-ion collisions do differ from the Gubser solution in meaningful ways.
The equation of state of nuclear matter~\cite{Borsanyi:2013bia, HotQCD:2014kol} is known to show significant deviations from conformality for temperatures between 100 to 400~MeV, with $c_s^2$ reaching a minimum of $c_s^2\approx 1/7$ around $T=170$~MeV; this slower speed of sound leads to a larger volume of plasma ($dV_{\perp}/dT$) at low temperature, which increases the number of photons produced there, and thus increases the potential impact that their Doppler shift can have on the total thermal photon energy spectrum. Moreover, the effect of viscosity has not been investigated in this work, and it also leads to changes in the volume distribution of plasma, as well as changes in the flow velocity profile and the photon emission rate itself.\footnote{Although the effect of bulk viscosity cannot be studied in any Gubser solution, the effect of shear viscosity could be studied in the future, for a constant shear viscosity over entropy density ratio $\eta/s=1/4\pi$.} A final difference with realistic simulations of heavy-ion collisions is that the Gubser solution does not take into account transverse flow resulting from fluctuations on scales smaller than the size of the nuclei~\cite{Chatterjee:2011dw,Dion:2011pp, Chatterjee:2012dn, Chatterjee:2013naa}. These fluctuations on nucleonic and subnucleonic scales will lead to a different and less structured pattern of transverse flow at early times, which will likely enhance the effect of Doppler shifts on high-energy photons. Some~\cite{vanHees:2011vb} or all~\cite{Shen:2013vja} of these effects are included in numerical studies of the transverse Doppler shift  and its effect on the photon spectrum. In practice, one also needs to consider other sources of photons, in particular prompt photons~\cite{Arleo:2011gc} which are rarely subtracted from photon measurements, unlike decay photons~\cite{David:2019wpt}. Prompt photons compete with thermal photons already in the $3$--$4$~GeV energy range.

Nevertheless, studying the simpler inviscid Gubser hydrodynamics made it possible to derive an exact expression for the transverse volume of the plasma as a function of temperature.
We further obtained a simple approximate formula, Eq.~\ref{eq:thermal_photon_spectrum_gubser_approx}, for the production of high-energy photons from a Gubser hydrodynamics profile with $q \tau_0 \ll 1$. We used this result to relate (i) the inverse exponential slope $T_{\textrm{eff}}$ of the photon energy spectrum to (ii) the actual maximum temperature of the plasma $T_0$. 
We found that the effective temperature $T_{\textrm{eff}}$ (Eq.~\ref{eq:Teff}) inevitably depends  on the photon energy range where $T_{\textrm{eff}}$ is fitted 
which can lead to large differences between $T_0$ and $T_{\textrm{eff}}$ when the latter is extracted from the lower-energy part of the photon spectrum.
Finally, as emphasized above, the results from Gubser hydrodynamics provided a simple example that illustrated the difference between the large \emph{local} effect of the Doppler shift on emitted photons and the smaller \emph{global} effect of the Doppler shift on the final energy spectrum of thermal photons.


\acknowledgements I thank Charles Gale and Berndt M\"uller for their valuable feedback on this manuscript, and Bj\"orn Schenke for his valuable feedback and his suggestion to add Figure~\ref{fig:Teff_vs_T0}. I am also grateful to the students of my fluid dynamics class for their perceptive questions about the Gubser solutions to relativistic fluid dynamics.

	\bibliographystyle{apsrev4-2}
	\bibliography{biblio}
	
\end{document}